\documentclass[aps,pre,twocolumn,showpacs,superscriptaddress]{revtex4}

\usepackage[latin1]{inputenc}
\usepackage[english]{babel}
\usepackage{bm,graphicx,graphics,amsmath,amssymb,epsfig,color}

\usepackage{dcolumn}
\newcommand{\be}{\begin{equation}}
\newcommand{\ee}{\end{equation}}
\newcommand{\bea}{\begin{eqnarray}}
\newcommand{\eea}{\end{eqnarray}}
\newcommand{\bse}{\begin{subequations}}
\newcommand{\ese}{\end{subequations}}
\newcommand{\comment}[1]{}

\begin{document}

\title{Heat transport in oscillator chains with long-range interactions \\ coupled to thermal reservoirs}
\author{Stefano Iubini}
\email{stefano.iubini@unifi.it}
\affiliation{ Dipartimento di Fisica e Astronomia and CSDC, Universit\`a di Firenze, via G. Sansone 1 I-50019, Sesto Fiorentino, Italy}
\affiliation{Istituto Nazionale di Fisica Nucleare, Sezione di Firenze, via G. Sansone 1 I-50019, Sesto Fiorentino, Italy}
\author{Pierfrancesco Di Cintio}
\affiliation{Consiglio Nazionale delle Ricerche, Istituto di Fisica Applicata ``Nello Carrara'' via Madonna del piano 10, I-50019 Sesto Fiorentino, Italy}
\affiliation{Istituto Nazionale di Fisica Nucleare, Sezione di Firenze, via G. Sansone 1 I-50019, Sesto Fiorentino, Italy}
\author{Stefano Lepri}
\affiliation{Consiglio Nazionale delle Ricerche, Istituto dei Sistemi Complessi, Via Madonna del Piano 10 I-50019 Sesto Fiorentino, Italy} 
\affiliation{Istituto Nazionale di Fisica Nucleare, Sezione di Firenze, via G. Sansone 1 I-50019, Sesto Fiorentino, Italy}
\author{Roberto Livi}
\affiliation{ Dipartimento di Fisica e Astronomia and CSDC, Universit\`a di Firenze, via G. Sansone 1 I-50019, Sesto Fiorentino, Italy}
\affiliation{Istituto Nazionale di Fisica Nucleare, Sezione di Firenze, via G. Sansone 1 I-50019, Sesto Fiorentino, Italy}
\affiliation{Consiglio Nazionale delle Ricerche, Istituto dei Sistemi Complessi, Via Madonna del Piano 10 I-50019 Sesto Fiorentino, Italy} 
\author{Lapo Casetti}
\affiliation{ Dipartimento di Fisica e Astronomia and CSDC, Universit\`a di Firenze, via G. Sansone 1 I-50019, Sesto Fiorentino, Italy}
\affiliation{Istituto Nazionale di Fisica Nucleare, Sezione di Firenze, via G. Sansone 1 I-50019, Sesto Fiorentino, Italy}
\affiliation{INAF - Osservatorio Astrofisico di Arcetri, largo Enrico Fermi 5, I-50125 Firenze, Italy}

\begin{abstract}
We investigate thermal conduction in arrays of long-range interacting rotors and Fermi-Pasta-Ulam (FPU) oscillators coupled to two reservoirs  at different temperatures. The strength of the interaction between two lattice sites decays as a power $\alpha$ of the inverse of their distance. 
We point out the necessity of distinguishing between energy flows \textit{towards/from the reservoirs}  
and those \textit{within the system}.
We show that energy flow between the reservoirs occurs via a \textit{direct transfer} induced by 
long-range couplings and a diffusive process through the chain.
To this aim, we introduce a decomposition of the steady-state 
heat current that explicitly accounts for such direct transfer of energy between the reservoir. 
For $0\leq \alpha<1$, the direct transfer term dominates, meaning that
the system can be effectively described as a set of oscillators 
each interacting with the thermal baths. Also, the heat
current exchanged with the reservoirs depends on the size of the 
thermalised regions: in the case in which such size is proportional 
to the system size $N$, the stationary current is independent on $N$.
For $\alpha > 1$, heat transport mostly occurs through diffusion along the chain:
for the rotors transport is normal, while for FPU the data are compatible
with an anomalous diffusion, possibly with an $\alpha$ -dependent characteristic exponent.
\end{abstract}


\maketitle

\section{Introduction}

The combination of various ingredients such as nonlinearity, reduced 
dimensionality, disorder and topology,
may yield quite a complex scenario of transport properties in many-body systems. In a general 
perspective, statistical mechanics should provide us the tools for investigating this problem
when a physical system is driven out of equilibrium by some external non-conservative force 
 or when a gradient is imposed by external reservoirs, exchanging energy, momentum and mass
 with the system. In particular, one is  interested in investigating stationary conducting states,
characterized by a minimal entropy production rate, whose physical manifestation is the presence of
stationary currents flowing through the system.  
  
Within this vast, interdisciplinary research domain, insight into   
universal features can be achieved by  studying simple paradigmatic models. 
Among them, systems of 
classical coupled oscillators are of particular interest, as they represent
a large variety of different physical problems, like atomic vibrations
in crystals and molecules or field modes in optics and acoustics.  

For nonlinear, low-dimensional lattice models there is currently a detailed understanding  
of their anomalous transport properties \cite{LLP03,Basile08,DHARREV,Lepri2016} leading to 
the breakdown of the classical Fourier law, typically  due to a superdiffusive transport mechanism. 

On the other hand, transport properties in long-range interacting systems has received some
attention only recently  \cite{avila2015length,Olivares16,bagchi2017thermal}. 
The problem is certainly far from trivial, because it is well known that these systems may relax 
to long-living  metastable states and exhibit anomalous diffusion of energy \cite{Bouchet2010,RuffoRev,Campa2014}.
Unusual effects like the lack of thermalization upon interaction with a single 
external bath \cite{deBuyl2013} or the presence, in isolated systems, of non-isothermal inhomogeneous
stationary states where the density and the temperature are anticorrelated 
\cite{2015PhRvE..92b0101T,gupta2016surprises} have been observed.
Another unusual feature is the fact that  
perturbations may propagate with infinite velocities making this class of systems 
qualitatively different from their short-ranged counterparts \cite{Torcini1997,Metivier2014}.

What should we expect when two or more external reservoirs drive a long-range interacting 
system in out-of-equilibrium conditions?
Some peculiar features, pointed out in a recent paper by Avila et al. \cite{avila2015length}
have been confirmed and more systematically investigated in a long-range coupled rotor
chain  \cite{Olivares16}. This model is quite interesting, because for nearest-neighbor (i.e. fully
localized) interactions it exhibits normal transport properties \cite{Giardina99,gendelman2000}, at variance with most models 
of nonlinear chains, that have been found to exhibit a diverging heat conductivity $\kappa$ with the system size
$N$ \cite{LLP03,Basile08,DHARREV,Lepri2016}. 
The main outcome reported in  \cite{Olivares16}  is that Fourier law is recovered only for sufficiently short-range 
interactions, while the bulk  conductivity seems to vanish in the mean-field limit. 

An analogous investigation has been
performed for the Fermi-Pasta-Ulam  (FPU) $\beta$-chain with long-range interactions \cite{bagchi2017thermal}. The main result in this case is that  $\kappa$ exhibits a power-law divergence with the system size  $N$ for any value
of the long-range exponent. Surprisingly enough  the divergence exponent of $\kappa$ does not depend monotonously on  
the range exponent. Moreover, for the range exponent equal to 2, it
takes a peculiar value close to 1, which seems to indicate a sort of
ballistic transport regime, as observed in integrable models, such as the harmonic chain 
\cite{RLL67}, or the Toda lattice \cite{Toda79,Kundu2016}.

In this paper we analyze similar long-range models and we introduce a suitable setup for clarifying the  basic
mechanisms underlying heat transport  processes (see Sec.~\ref{sec2}). The proper observables for
characterizing the non equilibrium stationary
regime in a system with long-range interaction are introduced in Sec.~\ref{sec3}, while in Sec.~\ref{sec4} we present the results obtained for the rotor chain.
The interpretation of the transport mechanism is outlined in terms of a two-component stationary heat
current (see Sec.~\ref{sec5}).
In Sec.~\ref{sec6} we extend the same kind of analysis to the $\beta$-FPU model. Conclusions and perspectives 
are summarized in Sec.~\ref{sec7}.

\section{The setup}
\label{sec2}
{\it Long-range Hamiltonian models -}
We consider a chain of $N$ anharmonic oscillators whose dynamics is governed by the Hamiltonian
\be
\label{eq:hamilt}
H = \sum_{i=1}^N \frac{p_i^2}{2m} + \sum_{i=1}^{N} \sum_{i\neq j}^N V(q_i,q_j) \, ,
\ee
where $q_i$ and $p_i$ are canonically conjugated variables, $V$ is a two-body interaction potential
and $m$ is the mass of each oscillator.
In order to have a direct control on the range of the interaction, we specialize $V$ to the following form
\be
\label{eq:int_pot}
V(q_i,q_j) = \frac{1}{2N_0(\alpha)} \frac{v(q_i-q_j)}{|i-j|^\alpha}  \, ,
\ee
where $\alpha$ is the range parameter and $N_0(\alpha)$ is a generalized Kac prescription, which
guarantees the extensivity of Hamiltonian (\ref{eq:hamilt}) in the thermodynamic limit. 
We have also assumed that the interaction between two sites $i$ and $j$ depends only on
the relative displacements $q_i-q_j$ through the function $v(q_i-q_j)$. 
For the potential in Eq. (\ref{eq:int_pot}), $N_0(\alpha)$ is defined as
\be
\label{eq:kac}
N_0(\alpha) = \frac{1}{N} \sum_{i=1}^N\sum_{j\neq i}^N \frac{1}{|i-j|^\alpha} \, .
\ee
Note that for $\alpha=0$ (mean field interaction) $N_0(0)=N-1$ reproduces  the standard Kac prescription, while
for $\alpha=+\infty$ (nearest-neighbour interaction) one has $N_0(+\infty)=1$. Overall, for any fixed size $N$,
$N_0(\alpha)$ is a monotonic decreasing function of $\alpha$, while  the interval $0\leq\alpha\leq 1$ identifies the
parameter region corresponding to  non-additive interactions in one dimension~\cite{RuffoRev}.


In this paper we are going to compare two different interaction functions $v(q_i-q_j)$, 
namely the one of the rotor-chain \cite{Tamarit2000} 
(sometimes called the $\alpha$-XY model or $\alpha$-HMF \cite{vanderberg2010}) 
\be
\label{eq:intXY}
v(q_i-q_j) =1-\cos(q_i-q_j)
\ee
and the one of the FPU-$\beta$ chain 
\be
\label{eq:intFPU}
v(q_i-q_j)= \frac{1}{2}(q_i-q_j)^2+\frac{1}{4}(q_i-q_j)^4 \, .
\ee
For the sake of simplicity we have eliminated any explicit dependence on any coupling parameter,
that have been set to unit. In fact, this choice simplifies  numerical studies, while 
it is not a prejudice of generality, since physical scales can be recovered by suitable
rescaling of time and of the amplitudes of the canonically conjugated variables $q_i$ and $p_i$.
For what concerns boundary conditions, they will be specified throughout the paper. Here we 
just want to point out that in a long-range interacting system driven in an out-of-equilibrium
stationary state
the choice of the boundary conditions deserves some attention.


In the limit $\alpha=\infty$, the above models reduce to their nearest-neighbor versions,
whose transport properties have been studied 
in great detail in the last two decades, see Refs.~\cite{Giardina99,gendelman2000,Yang2005,Iubini2016}
and \cite{LLP97,Lepri05,Wang2011}, respectively.
For finite $\alpha$, transport properties of the rotor model (\ref{eq:intXY}) have 
been studied recently in Ref. \cite{Olivares16}. We remark that the FPU model considered
in Ref.~\cite{bagchi2017thermal} differs from (\ref{eq:intFPU}) as the long-range 
interaction is only in the nonlinear term, while here we consider the case in which the 
long-range terms are present already in the harmonic limit. Other variants of the 
model where nonlinearity is local have been considered in the literature 
\cite{Miloshevich2015,Miloshevich2017}.



\begin{figure}[ht]
 \includegraphics[width=0.48\textwidth,clip]{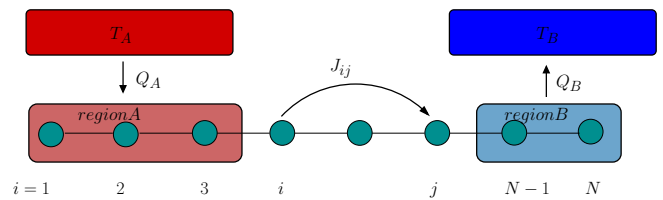}
 \caption{
  A schematic representation of the nonequilibrium setup.  Regions $A$ and $B$ (rounded boxes) identify the oscillators that
 interact with heat baths at temperature $T_A$ and $T_B$ respectively, with $l_A=3$ and $l_B=2$.
 Heat currents from the reservoirs to the system are denoted
 with $Q_A$ and $Q_B$ and those flowing from two generic sites of the chain by $J_{ij}$
 respectively.
 }
 \label{fig:chain}
\end{figure}

{\it Heat transport in a temperature gradient -}
Nonequilibrium stationary states (NESS's) are studied  by
attaching reservoirs at
temperature $T_A$ and $T_B$ to the particles in two different regions of the system, A and B, extending 
over $l_A$ and $l_B$ lattice sites, respectively (e.g., see Fig \ref{fig:chain}).

The reservoir dynamics here implemented corresponds to the so-called Maxwellian heat bath \cite{LLP03}:
thermalized system particles interact by a sequence of elastic collisions with particles of mass
$m_{gas}$, selected from a one-dimensional ideal gas in equilibrium at temperature $T$.
The corresponding numerical algorithm can be simplified by adopting the condition  $m_{gas}=m$,  
so that  each collision event reduces  to exchanging the momentum $p_i$ of the system particle 
with the momentum of the selected gas particle.
The  sequence of time intervals $\tau$ between collisions is obtained by a Poissonian 
distribution
\begin{equation}
\label{poisson}
 P(\tau) = \gamma\exp(-\gamma \tau) \, ,
\end{equation}
where $\gamma$ is the strength of the coupling with the Maxwellian heat bath. 
For the numerical simulations presented in this paper, we have chosen $\gamma=1$ and $m=1$.
Denoting with $\tau_i^A$ the  sequence of the Poisson-distributed time intervals, labeled 
by the integer index $i = 1, 2, \dots$, in region $A$,
the $n$-th collision event occurs at time $t_n^A = \sum_{i=1}^n \tau_i^A$. The same definition  holds for collision 
timing in region $B$, just by changing label $A$ with $B$.  If the interaction regions with the reservoirs 
involve more than one lattice site (i.e. $l_{A}>1$ or $l_B>1$), independent collisions are generated on each site.
 
Note that during the time between two consecutive collisions the dynamics of the system is Hamiltonian. 
We have numerically integrated the equations of motion $\dot p_i = -\partial H / \partial q_i$ and $\dot q_i = \partial H / \partial p_i$
using a symplectic 4-th order  McLachlan-Atela algorithm \cite{mclachlan1992accuracy}, with time step $\delta t =0.05$.  This choice guarantees
a sufficiently accurate sampling of trajectories in all the conditions  reported throughout this paper.
More precisely, the simulation protocol here adopted amounts to the following steps. First, the system
is evolved to a thermal equilibrium state at temperature $T=(T_A+T_B)/2$. 
Then, it is further evolved only in the presence  of a temperature gradient, imposed by the heat baths at temperature $T_A$ and $T_B$,  acting in regions $A$ and $B$. This evolution lasts over  a time interval $t_s$, 
long enough for observing a NESS, characterized by a stationary entropy production
rate, together with stationary currents and temperature profiles.  In general,  $t_s$ depends on the system size $N$
and on the range parameter $\alpha$: in the whole range of $\alpha$ and  for the largest systems sizes explored in this paper we have checked that  
$t_s \sim {\mathcal O} (10^7)$ time units is sufficient for obtaining an effective sampling of the NESS's. 
 
Assuming that local equilibrium conditions set in the NESS, the temperature profile $T_i$ is computed 
as the time average of the local momentum fluctuations with $m=1$, i.e.
\be
T_i =  \langle p_i^2 \rangle -\langle p_i\rangle^2  \, .
\ee
In order to obtain a clear understanding of energy transport mechanisms in long-range models it is 
useful introducing suitable definitions of heat currents, that we are going to illustrate in the next section.


\section{heat currents }
\label{sec3}
In the out-of-equilibrium setup sketched in Fig.~\ref{fig:chain}, stationary heat currents depend on the boundary conditions 
imposed on the thermalized regions $A$ and $B$, such as the boundary temperatures $T_A$ and $T_B$
and the bath coupling strength $\gamma$. While this scenario is well understood for short-range systems~\cite{LLP03},
several unusual properties arise when long-range interactions are considered. 
In order to analyze the NESS  generated by a given  thermal gradient,
it is first necessary to define the heat current flowing  from region $A$ to region $B$.  
This task can be  accomplished by computing  the heat currents, $Q_A$ and $Q_B$, exchanged  by the system  
with the reservoirs at temperatures $T_A$ and $T_B$.
This  amounts to measure the time average of the variation of kinetic energy of the particles
in contact with the reservoirs in regions $A$ and $B$.
More precisely,  
we define
\be
\label{flux}
Q_A=\lim_{t\to \infty}\frac{1}{t} \sum_{t_n^A\le t}\,\,\sum_{i \in A}  \delta K_i(t_n^A) \, ,
\ee
where   $\delta K_i(t_n^A)$ is the variation of local kinetic energy produced by the collision 
of the Maxwellian heat bath with the  particle in  $i \in A$ at time $t_n^A$.
The definition of $Q_B$ is readily obtained by changing label $A$ with $B$ in (\ref{flux}). 

When a NESS is reached, the energy balance condition reads
\be
\label{eq:station}
Q_A = - Q_B \, ,
\ee
In what follows, rather than $Q_A$ and $Q_B$ separately, we prefer to consider the
average total heat current
\be
\label{eq:avstation}
Q = (Q_A - Q_B)/2 \, ,
\ee
because it incorporates the fluctuations originated by both heat reservoirs.
  
In addition to this global observable, it is worth analyzing the average local
heat current $J_{ij}$ from site $i$ to site $j$, i.e. 
\be
\label{locflux}
J_{ij} = \left \langle \frac{1}{2} F_{ij} (p_i+p_j)  \right\rangle \, ,
\ee
where $F_{ij}=-\partial V(q_i,q_j)/\partial{q_i} $ is the force exerted on the particle at site $i$ by the
one at site $j$. This definition stems from the condition of local energy conservation ~\cite{LLP03}.
Note that, due to the symmetry of $V(q_i,q_j)$, $J_{ij}$ is antisymmetric under the exchange of indices $i$ and $j$.

The current-matrix $\hat J$, whose entries are the average local currents $J_{ij}$ defined
by Eq.~(\ref{locflux}), provides information on the 
relevant {\it internal} channels of the system, through which 
heat can flow. It should be clearly distinguished from the flow towards the 
the reservoirs, $Q_A$ and $Q_B$  and to this aim we use for it a different symbol.
As shown in ~\cite{liu2007heat,liu2010heat}, the observables defined in Eqs. (\ref{flux}) and (\ref{locflux}) 
have proved to be useful for a clear understanding of heat transport in 
generic oscillator networks.
In what follows, we are going to discuss how  these observables 
depend on  the range parameter $\alpha$.

One of the key features displayed by systems with long-range interactions is the lack of additivity~\cite{RuffoRev}:
 even in the thermodynamic limit, it is impossible to decompose an initial system at thermodynamic equilibrium into two
effectively noninteracting subsystems. Equivalently, when one puts in contact two identical and independent long-range subsystems at thermal equilibrium,
the whole system may display properties that do not correspond to the original equilibrium state.
Therefore, when dealing with thermal conduction problems, a natural question is whether a similar indecomposability
holds in the nonequilibrium setup shown in  Fig.~\ref{fig:chain}, where one has implicitly identified a bulk system and two thermalized leads.

To better investigate this point, we have considered two different coupling schemes
that specify the system-reservoir interaction.
In the first scheme ({\sl extensive coupling}) we introduce a finite fraction of  thermalized  sites in both regions
A and  B, i.e.  we assume that
 $l_A$ and $l_B$ increase linearly  with the system size $N$.
In the second scheme ({\sl finite coupling}), we assume that $l_A$ and $l_B$ are fixed, independently of $N$.

\section{the rotor chain}
\label{sec4}
In a first series of simulations, we have studied the NESS's of the rotor chain
(\ref{eq:intXY}) for different
values of the range parameter $\alpha$ and for different thermal coupling schemes. 

\begin{figure}[ht]
\includegraphics[width=0.45\textwidth,clip]{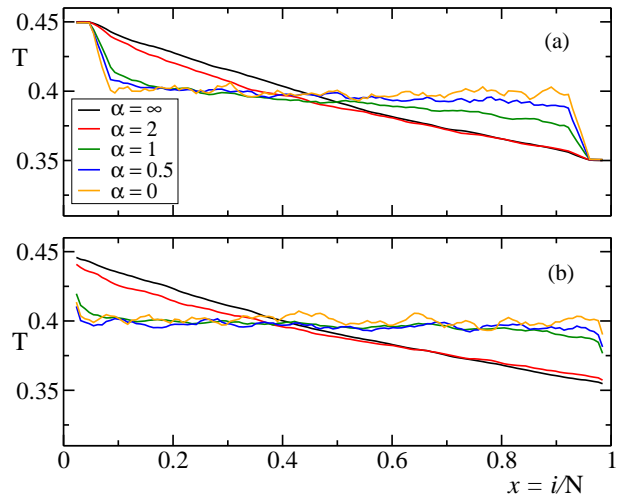}
 \caption{Stationary XY temperature profiles for the setup in Fig.~\ref{fig:XYfluxscal_dc} and $N=128$.
 Panel (a) refers to the  extensive-coupling case with $l_A=l_B=N/16$, while panel (b) shows the profiles for the 
 finite-coupling case,
 with $l_A=l_B=1$. Simulations have been performed by setting  $t_s=10^7$. All the other parameters  
 are the same of  Fig.~\ref{fig:XYfluxscal_dc}.
 \label{fig:XYTprof}
 }
\end{figure}

In Fig.~\ref{fig:XYTprof} we show the stationary temperature profiles  
for a chain in contact with two reservoirs at temperature $T_A= 0.45$ and $T_B = 0.35$
and fixed boundary conditions. The resulting NESS's  have been studied  for both the extensive 
coupling (see panel (a)) and for the finite coupling  (see panel (b)). In agreement with previous
observations obtained within the finite coupling scheme~\cite{Olivares16}, we find that the temperature
profiles are almost flat for $\alpha\leq 1$, while they  are linear for $\alpha>1$. 
Moreover, the profiles do not display any relevant dependence on  the choice
of the coupling scheme.

\begin{figure}[ht]
 \includegraphics[width=0.45\textwidth,clip]{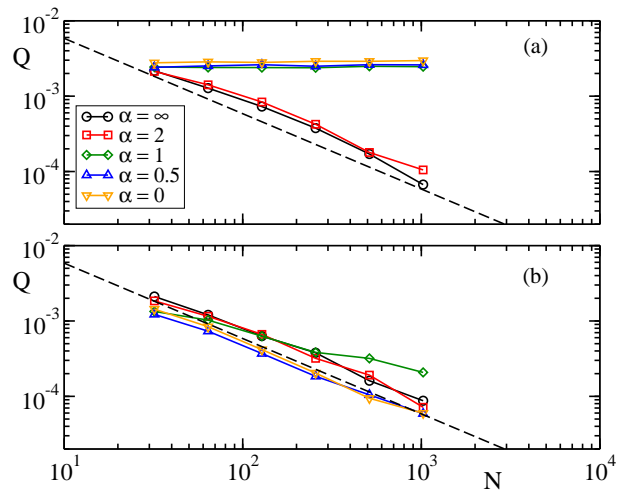}
 \caption{Scaling of the stationary heat current $Q$ in the XY chain  as a 
 function of the system size $N$ and for different range parameters $\alpha$.
 Panel (a) refers to the extensive thermal coupling, with a size $l=N/16$
 for both the thermalized regions A and B, while panel (b) shows the results for 
finite  thermal coupling with $l_A=l_B=1$. In both panels,
 the black dashed line refers to  the power-law scaling $N^{-1}$.
 Simulations have been performed by setting $T_A=0.45$, $T_B=0.35$ and $t_s=10^6$. 
 }
 \label{fig:XYfluxscal_dc}
\end{figure}
Quite a different behavior is observed when one analyzes the total average heat current  $Q$
exchanged with the external reservoirs
(see Eq.~(\ref{eq:avstation})).
Let us discuss this point starting
from the  case $\alpha=\infty$, that refers to a rotor chain with nearest-neighbor interactions.
In this case, normal heat 
conduction was observed as a consequence of the periodicity of the interaction potential~\cite{gendelman2000,Giardina99}. 
In fact, we recover the  characteristic diffusive scaling
$Q\sim N^{-1}$ for both choices of the thermal coupling (see the black circles in
both panels of  Fig.~\ref{fig:XYfluxscal_dc}). Such a scaling is maintained also when
the range parameter is reduced down to $\alpha > 1$, i.e. when additivity is preserved: 
in both panels of  Fig.~\ref{fig:XYfluxscal_dc} we have just reported
the cases $\alpha = 1$ and  2. 
These results indicate that heat transport occurs essentially via energy diffusion through the bulk
of the system for $1 < \alpha < \infty$ and independently of the extensive- or 
finite-coupling scheme.

Conversely, for $\alpha < 1$ the current 
$Q$ displays two different scalings,  depending on the choice of the coupling scheme. More precisely, 
for the extensive-coupling case (see panel (a) in Fig.~\ref{fig:XYfluxscal_dc}) $Q$ is essentially constant 
with $N$, while for 
the finite-coupling case (see panel (b) in Fig.~\ref{fig:XYfluxscal_dc}) $Q$ scales again as $N^{-1}$.
We have checked that this scenario is not altered when dealing with a chain with periodic boundary
conditions: in this case the chain turns  into a ring of length $2 \times N$, where also $l_{A,B} \rightarrow
2 \times  l_{A,B}$, and the maximum distance between a pair of particles is $N$. 

\begin{figure*}[ht]
\begin{center}
\includegraphics[width=0.2\textwidth,clip]{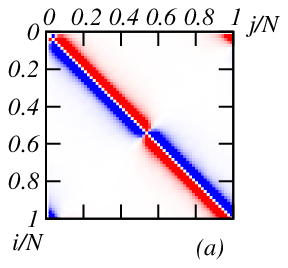}
\includegraphics[width=0.2\textwidth,clip]{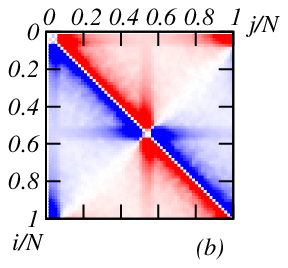}
  \includegraphics[width=0.2\textwidth,clip]{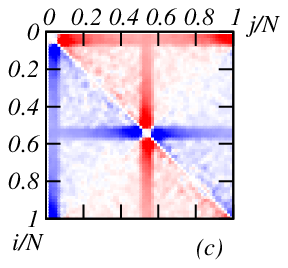}
   \includegraphics[width=0.255\textwidth,clip]{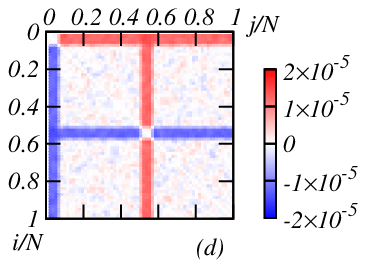}
   \includegraphics[width=0.45\textwidth,clip]{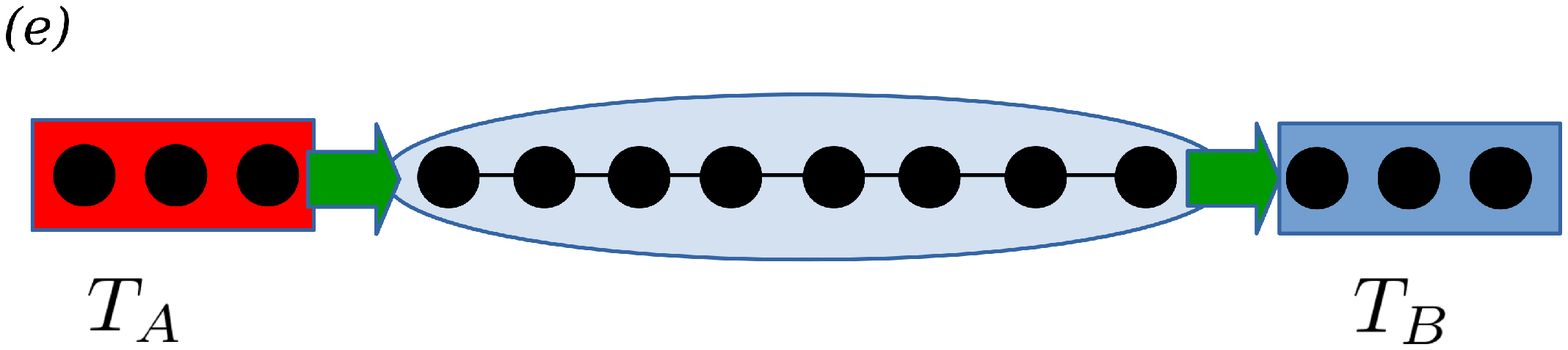}
   \includegraphics[width=0.23\textwidth,clip]{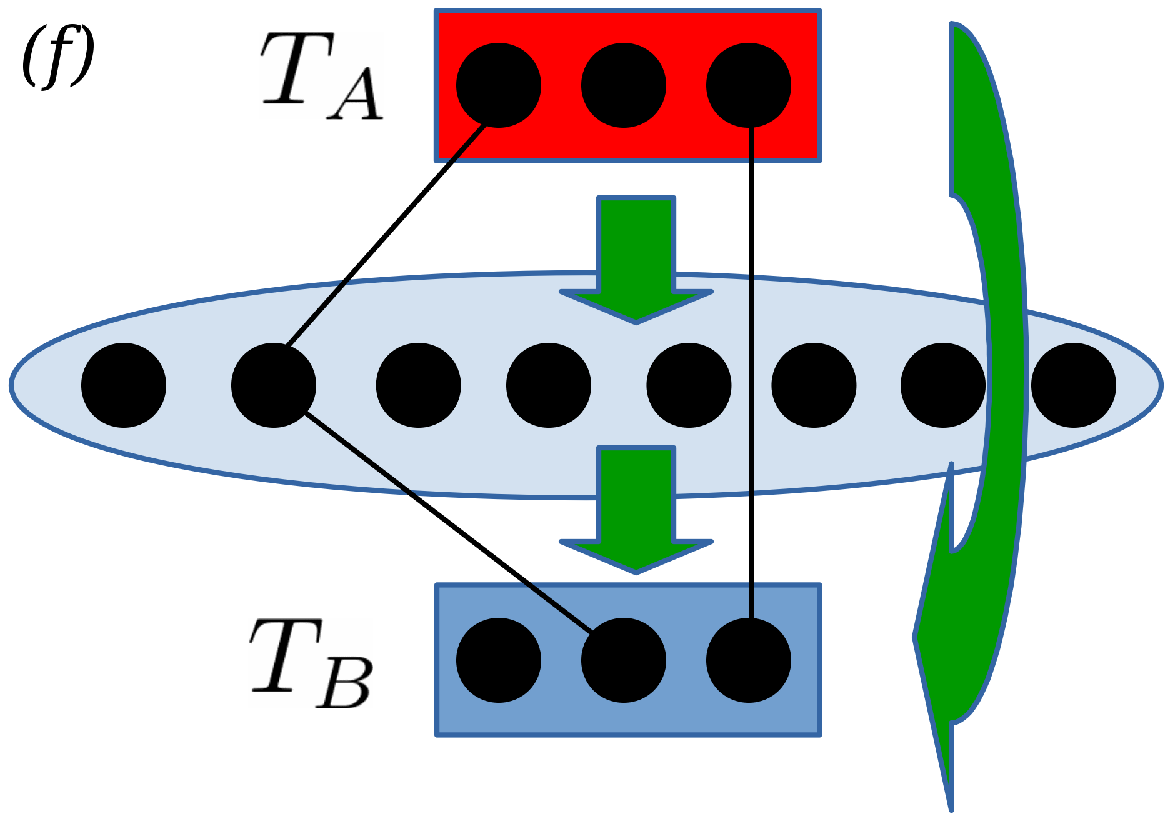}
\end{center}
\caption{Stationary current-matrix $\hat J$, represented as a function of the intensive variables $i/N$ and $j/N$ 
for a rotor chain with $2N= 64$ (periodic boundary conditions) and  $t_s=5\times10^7$. 
The chain is in contact with two reservoirs at temperature $T_A=0.45$ and $T_B=0.35$ with
extensive coupling, i.e. $l_A=l_B=N/16$. The color code
points out the main channels through which heat flows.
Panels (a), (b), (c) and (d) refer to $\alpha=2,\,1,\,0.5,\,0$, respectively.  
Panels (e) and (f) are pictorial representations of the heat transfer process for $\alpha=\infty$ and
$\alpha=0$, respectively. Thermalized sites are contained in the rectangular boxes while the bulk system
is represented in the ellipse. Black lines identify the relevant transport channels. }
 \label{fig:FMATXY}
\end{figure*}

It is important to point out that both of the situations observed for
$\alpha < 1$ are originated by the
presence of effective ``short circuits'', that  allow  for
a direct energy transfer between the thermalized leads.  In particular, the $N^{-1}$-scaling 
of $Q$ observed in the finite-coupling case has nothing to do with energy diffusion, as we
discuss in Sec.\ref{sec5}. An intuitive argument  in the mean-field limit  $(\alpha=0)$ is that
any pair of thermalized sites $i\in A$ and $j\in B$  can
interact directly through a long-range force $F_{ij}$. However, it is less evident to figure out how
preferred channels of heat transport organize, when $N$ mutually interacting oscillators are considered 
and, more importantly, how  they depend on the range parameter $\alpha$. 
In order to spotlight these peculiar energy pathways, we have computed the stationary
current-matrix $\hat J$ for different values of $\alpha$. The results are shown in Fig.~\ref{fig:FMATXY} for
a system with $2 N=64$ sites (periodic boundary conditions) in the presence of the extensive thermal coupling.
It is useful to recall that for a rotor chain with pure nearest-neighbor interactions $(\alpha=\infty)$,
the local conservation
of energy implies~\footnote{Here we assume for simplicity that the sites $i$ and $j$ are not in contact with the external reservoirs.} 
\be
\label{eq:fluxNN}
|J_{ij}|=c(\delta_{i, j+1}+\delta_{i, j-1}) \quad ,
\ee
 where $c=\mathcal{O}(N^{-1})$
~\cite{Giardina99,gendelman2000}.
As a result $\hat J$ is a tridiagonal matrix with vanishing elements on the diagonal (data not shown) 
and it reflects the existence of a channel through which energy is transported, see  panel (e).
For $\alpha=2$ (panel (a)) $\hat J$ maintains essentially the same properties of the nearest-neighbor case, although the nonvanishing 
elements are now  smeared over a broader region around the diagonal.
Upon reducing the range parameter from $\alpha=2$ to $\alpha=0$, this structure progressively disappears
in favor of a new pattern which is organized over horizontal and vertical bands, that 
involve the dominant contribution coming from thermalized sites 
(panels (b), (c), (d)). For $\alpha=1$ a coexistence of the two patterns is observed, while for $\alpha=0$ the original
diffusive pattern discussed in the nearest-neighbor  limit has completely disappeared.
A closer inspection to panel (d) reveals that the two reservoirs exchange energy either via a direct transfer
between two thermalized sites or via a mediated process, that involves a generic site that is not directly
thermalized. This mechanism is sketched in panel (f) and provides a simple argument to explain the presence
of flat temperature profiles close to $\alpha=0$, as shown in Fig.~\ref{fig:XYTprof}.
In practice, each lattice site in the bulk is effectively coupled both to the hot and to the cold reservoir.
As a result, its temperature settles to the average value $T=(T_A+T_B)/2$ independently of $N$, so that the average
heat current exchanged with any other site in the bulk is practically negligible. 

Altogether, by comparing panels (e) and (f) in Fig.~\ref{fig:FMATXY}, we conclude that in the mean-field
limit, $\alpha=0$, the system is organized as an array of oscillators connected {\it in parallel} to the reservoirs.

\section{Decomposition of stationary flow}
\label{sec5}

In this section, we clarify the relation between the geometrical reorganization of the
energy channels in the current-matrix $\hat J$ shown in Fig.~\ref{fig:FMATXY} and the scaling of the 
global heat current $Q$ (see Fig.~\ref{fig:XYfluxscal_dc}).

Before addressing the case of a generic $\alpha$, let us first discuss the two 
extreme cases of mean-field and short range interaction, respectively.
In the former case $\alpha=0$, the total heat current
$Q$ is constant for the extensive coupling and scales like $N^{-1}$ for the finite coupling.
These numerical results can be explained in terms of a two-site diffusion process between a thermalized site and a generic bulk site, whose temperature $T$ does not depend on $N$.
Indeed, let us evaluate how the current element $J_{ik}$, with $i\in A$ representing a generic site 
thermalized at $T_A$ (the donor) and $k $ a generic 
site in the bulk of the chain (the acceptor) at temperature $T$, scales with the system size $N$. Since $\mathcal{O}(N)$
equivalent  energy channels are available for the donor site $i$, an energy
fluctuation created on it has a probability $\mathcal{O}(N^{-1})$ to be transferred to the acceptor site $k$.
Moreover, due to the Kac prescription (see  Eq.~(\ref{eq:kac})), the interaction force $F_{i k}$ scales as
$N^{-1}$. As a result, the generic current element $J_{ik}$ scales as $N^{-2}$.
Since in any NESS the total current $Q_A$, defined in Eq.~(\ref{flux}), has to be equivalent also to the
following expression 
\be
\label{eq:JA}
Q_A=\sum_{i\in A,\,k} J_{i k} \, ,
\ee
we can conclude that $Q_A$ is constant for the extensive coupling and $Q_A \sim N^{-1}$ for 
the finite coupling,
in agreement with the numerical results.

We point out that the above mechanism does not depend on the details of the interaction potential $v(q_i-q_j)$. Indeed, a similar result is found for the long-range FPU chain (see Sec.~\ref{sec6}). In addition, we have verified that the patterns of the current-matrix $\hat J$, obtained  for $\alpha<1$, are stable with the system size $N$. 

In the short range limit $\alpha=\infty$, the above two-site (donor-acceptor) picture does 
not apply: for the case of the rotor chain transport is known to be normal 
\cite{Giardina99,gendelman2000} and energy transfer occurs through 
a standard diffusion process, going
through all the sites of the chain. In this case, the nonvanishing current elements $J_{i k}$  are proportional 
to the global temperature gradient $(T_A-T_B)/N$ and therefore $J_{i k}$ scales
as  $N^{-1}$.

Altogether, these considerations suggest that for any finite $\alpha$, the stationary heat current 
flowing through the chain is a combination  of  mean-field-like and local components. 
This is why it is worth describing the heat transport process in long-range models
in terms of a two-component flow. 
Let us make explicit the dependence of  the current-matrix on the range parameter, i.e. $\hat J(\alpha)$.
Again we want to focus our attention on the matrix elements $J_{ik}(\alpha)$, where $i \in A$ and $k\not\in A$.

We assume that for $N$ large, we can decompose $J_{ik}(\alpha)$ as follows
\be 
\label{eq:dec}
J_{ik}(\alpha) = \frac{1}{N^2}\,f^{(\alpha)}\left(\frac{k}{N}\right) +\frac{1}{N} g_{ik}^{(\alpha)} \, ,
\ee
where $f^{(\alpha)}(x)$ and $g_{ik}^{(\alpha)}$ are summable quantities, namely
\begin{eqnarray}
\sum_k f^{(\alpha)}\left(\frac{k}{N}\right) &=&  N \int_0^1 f^{(\alpha)}(x)\, dx = NF^{(\alpha)}\\
\sum_{i,k} g_{ik}^{(\alpha)} &=&  G^{(\alpha)} \, .
\end{eqnarray}
We assume also that 
$F^{(\alpha)}$ and $G^{(\alpha)}$ are finite quantities also in the thermodynamic limit. 
Upon this decomposition, the total heat current $Q_A$ defined in Eq.~(\ref{eq:JA}) can be
expressed in the form
\be
\label{eq:dec2}
Q_A(\alpha) = \frac{l_A}{N} F^{(\alpha)} +\frac{1}{N} G^{(\alpha)}\, ,
\ee
where the first term  accounts for the long-range contribution to heat transport, while
the second one accounts for the short-range diffusion process. Note that the factor $l_A$
in front of the first addendum is a consequence of the assumption that the quantities $f^{(\alpha)}$
and $ F^{(\alpha)}$ are independent of $i$, i.e. the long-range contribution from the thermalized
sites in region $A$ is the same from all of these sites.

\begin{figure*}[ht]
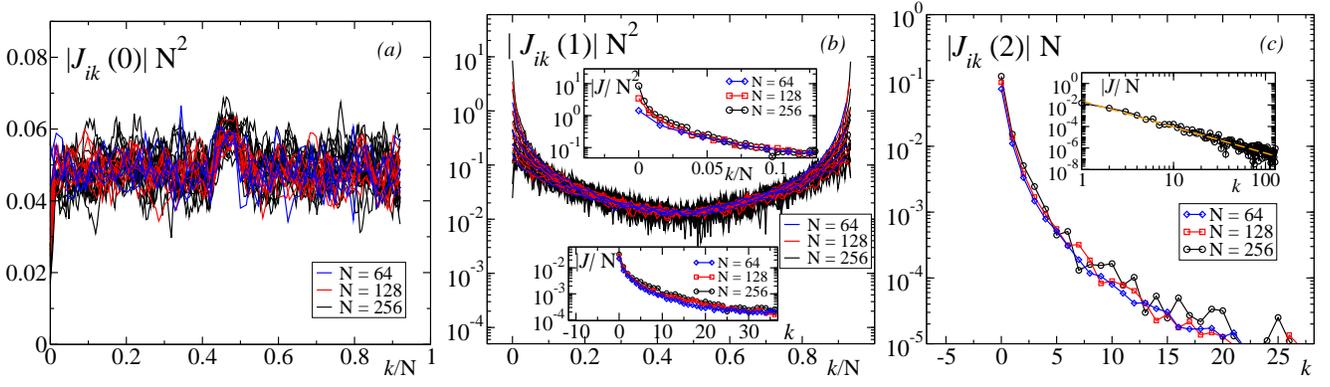

\begin{center}
\includegraphics[width=0.32\textwidth,clip]{{flux_components_h0.000_abs}.eps}
\includegraphics[width=0.32\textwidth,clip]{{flux_components_h1.000_abs}.eps}
\includegraphics[width=0.32\textwidth,clip]{{flux_components_h2.000_abs_v2}.eps}
\end{center}
\caption{Spatial profiles of the modulus of $J_{ik}{(\alpha)}$ in the rotor chain for different system size $N$ and extensive coupling $l_A = l_B=N/16$:
 $\alpha=0$ (panel (a)),  $\alpha=1$ (panel (b)) and $\alpha=2$ (panel (c)).
To obtain a cleaner evidence of the data collapse in  the lower and upper inset  of panel (b) and for  the profiles in panel (c)  we have  just 
reported $J_{-1 k}(\alpha)$. The setup for numerical simulations is the same adopted for Fig.~\ref{fig:FMATXY}.}
 \label{fig:F_comp}
\end{figure*}

The two contributions in Eq.~(\ref{eq:dec}) for different values of the range parameter $\alpha$  are  highlighted  
in Fig.~\ref{fig:F_comp}, that has been obtained for chains with periodic boundary conditions.
In particular, we show the current profiles $J_{ik}(\alpha)$ obtained from numerical simulations of the rotor chain with periodic boundary conditions and extensive coupling, $l_A = l_B = N/16$. 
In this setup,  Eq.~(\ref{eq:dec2})
predicts  that long-range contributions to the heat current are constant, while short-range ones scale as $N^{-1}$.
In order to separate visually the thermalized sites from the bulk sites, in these pictures we have
chosen a spatial reference frame where the site $i=0$ labels the first particle which is {\sl not} in contact with the 
heat bath at temperature
$T_A$.
Accordingly, the thermalized sites, $i\in A$,
have a negative space coordinate for any $N$.
For $\alpha=0$ (see panel (a) in Fig.~\ref{fig:F_comp}) we observe that the rescaled observable  $J_{ik}(0) \, N^2$ is essentially independent of $N$,
thus proving that  the only relevant term in Eq.~(\ref{eq:dec}) is due to $f^{(0)}\left(\frac{k}{N}\right)$, while $g_{ik}^{(0)}$ has to vanish.
The flat profile indicates that each thermalized site transfers energy uniformly to all the other sites of the bulk. 
For $\alpha=1$ (panel (b)), the rescaled observable $J_{ik}(1) \, N^2$ exhibits a good data collapse in  a macroscopic region $(\mathcal{O}(N))$
region around $k/N = 0.5$, where the reservoir at temperature $T_B$ is located. This indicates that the relevant term 
for large $N$ in Eq.~(\ref{eq:dec}) is due to $f^{(1)}\left(\frac{k}{N}\right)$. On the other hand, in a limited region close to $k=0$ we observe
a significative deviation from the data collapse (see the upper inset in  panel (b)). In fact, in the lower inset of panel (b) we show that the 
rescaled quantity $J_{ik}(1) \, N$ exhibits a definitely better data collapse, thus indicating that the main contribution in this region is due to $g_{ik}^{(1)}$.
In summary, for $\alpha = 1$ we are facing a situation where the two contributions appearing in Eq.~(\ref{eq:dec}) do not vanish: the first
one determines the scaling behavior of $Q_A(1)$ in the large $N$ limit  (i.e., it is constant), while the other simply affects corrections to
scaling in a limited region around $k=0$.
 Finally, for $\alpha=2$ (see panel (c)), we observe that a a good data collapse is obtain for the rescaled observable $J_{ik}(1) \, N$:  
 $g_{ik}^{(2)}$  provides the main contribution to Eq.~(\ref{eq:dec}) relevant,
while the long-range component $f^{(2)}\left(\frac{k}{N}\right) $ vanishes.
Moreover, the quantity $g_{ik}^{(2)}$ is found to decay  quite rapidly as a function of the distance from the reservoir at $T_A$. From a fit obtained for the chain with largest size analyzed in this paper ($N=256$), we can
estimate an asymptotic power-law decay as $g_{ik}^{(2)}\sim k^{-\rho}$ with $\rho\simeq 2.4$ (see the inset in panel (c)). This result confirms the summability of $g_{ik}^{(2)}$ over $k$, as required for the ansatz~(\ref{eq:dec}). We have also checked the summability of $g_{ik}^{(2)}$ over the reservoir indices $i$
(data not reported). 

We can conclude that this phenomenological description of the scaling of the total heat current $Q_A(\alpha)$
with the system size $N$ in terms of a two-component stationary flow is able to capture the basic mechanisms
that coexist  and eventually prevail onto each other in the long-range rotor chain. Pictorially, the overall information
is summarized in the form of the current matrix $\hat J (\alpha)$, shown in the upper panels of Fig.~\ref{fig:FMATXY}. 

\section{the FPU chain}
\label{sec6}
Here we extend our study of heat transport in long-range models to the
FPU-$\beta$ chain, whose interaction potential is given 
in Eq.~(\ref{eq:intFPU}). We report just the main results, without going through a detailed
description as for the rotor chain. 





\begin{figure}[htbp]
\includegraphics[width=0.45\textwidth,clip]{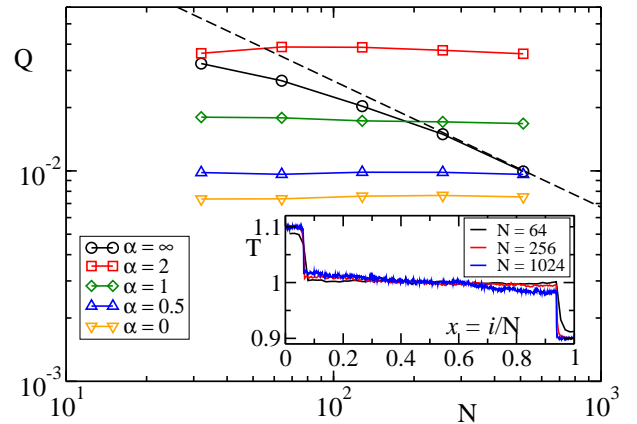}
 \caption{Scaling of the stationary heat current $Q$ of the FPU chain as a function of the system size $N$ for different range parameters $\alpha$. The black dashed line refers to a power-law $Q\sim N^{-\frac{3}{5}}$.  Simulations refer to an extensive thermal coupling with $l_A=l_B=N/16$ and  fixed boundary conditions. Nonequilibrium states were sampled by choosing $t_s=10^6$.
 The inset shows the temperature profiles  for $\alpha=2$ and  different chain lengths $N$. Note that the  horizontal axis reports  the rescaled spatial variable $x=n/N$.
 \label{fig:FPUfluxscal}
 }
\end{figure}
\begin{figure*}[ht]
\begin{center}
\includegraphics[width=0.2\textwidth,clip]{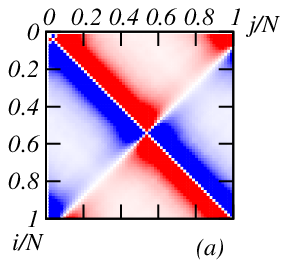}
\includegraphics[width=0.2\textwidth,clip]{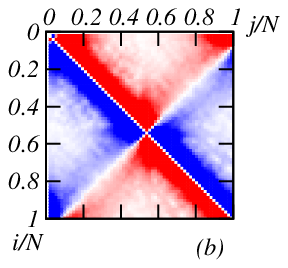}
\includegraphics[width=0.2\textwidth,clip]{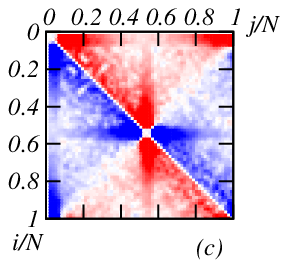}
\includegraphics[width=0.255\textwidth,clip]{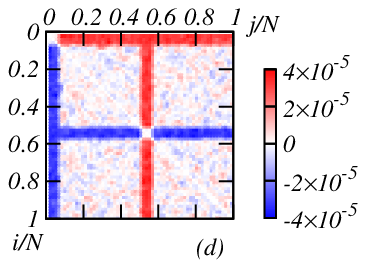}
\end{center}
\caption{Stationary current-matrix $\hat J$ as a function of the intensive variables $i/N$ and $j/N$ for a long-range FPU chain with $N=64$ and periodic boundary conditions and $t_s=10^7$.
The chain is in contact with two reservoirs at temperature $T_A=1.1$ and $T_B=0.9$ with coupling lengths  $l_A=l_B=N/16$.
Panels (a), (b), (c) and (d) refer to $\alpha=2,\,1,\,0.5,\,0$, respectively.  
}
 \label{fig:FMAT}
\end{figure*}

A first series of numerical simulations have been performed with the nonequilibrium setup described in Sec.~\ref{sec2} with $T_A=1.1$ and $T_B=0.9$.
Fig.~\ref{fig:FPUfluxscal} shows the scaling of the stationary total heat current $Q$
as a function of the system size $N$ for the extensive thermal coupling and fixed boundary conditions.  
We have verified that for $\alpha = \infty$ (i.e., nearest-neighbour interaction) numerics essentially
agrees with the expected anomalous scaling $Q \sim N^{- \frac{3}{5}}$ 
consistent with previous studies of the short-range case \cite{LLP2003,Pereverzev2003,Lukkarinen2008,Wang2011}. 
The bending of the
black circles with respect to the dashed line in Fig.~\ref{fig:FPUfluxscal} is due to finite size effects.
We want to stress that this scenario is  different from the rotor chain, which exhibits normal conductivity,
i.e. $Q \sim N^{- 1}$, in the nearest-neighbour limit. 
On the other hand,  for $\alpha \leq 1$ also for the FPU-$\beta$ chain we have evidence that 
$Q$ does not depend on the system size $N$, as shown
in Fig.~\ref{fig:FPUfluxscal} for the cases $\alpha = 0, 0.5$ and 1. The only difference with respect to
the rotor chain is that the value of $Q$ decreases more rapidly with $\alpha$. 
This result indicates that the basic mechanism of heat transport in this  non-additive regime is the same 
``parallelization'' process of  heat channels discussed in Fig.~\ref{fig:FMATXY} for the rotor chain. Indeed, the current-matrix $\hat J$,
( see panels (b),(c) and (d) in Fig.~\ref{fig:FMAT}) exhibits patterns that are very similar to 
those shown in Fig.~\ref{fig:FMATXY}, with a fully parallelized structure for $\alpha=0$. 

\begin{figure}[ht]
\includegraphics[width=0.45\textwidth,clip]{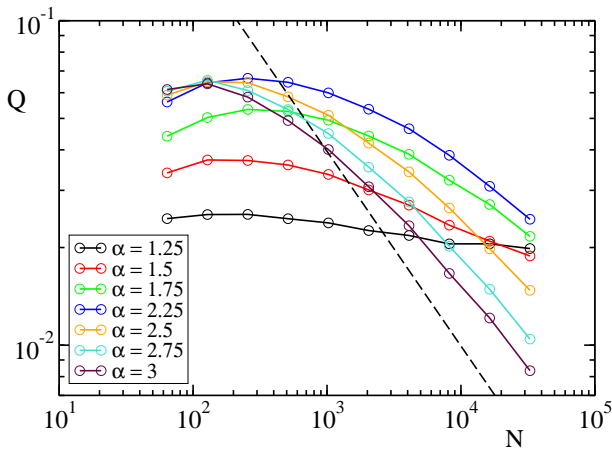}
 \caption{Scaling of the stationary heat current $Q$  of the FPU chain as a function of the system size $N$ for different values of the range
 parameter $\alpha>1$ and $t_s=5\times 10^6$. Extensive coupling and periodic boundary conditions have been adopted for these numerical simulations. 
  The black dashed line refers to a power-law $Q\sim N^{-\frac{3}{5}}$.
 \label{fig:FPUfluxscal2}
 }
\end{figure}

The scenario for $1 < \alpha < \infty$ is definitely richer than in the rotor chain.
The dependence of $Q$ on $N$ for various values of $\alpha$ in this range is
reported in Fig.~\ref{fig:FPUfluxscal2}.  On the basis of these results we can speculate that
the closer $\alpha$ is to 1, the more finite size effects prevent the possibility of recovering the anomalous scaling of the
nearest-neighbour case. In fact, all curves in this Figure exhibit a clear tendency to bend for
increasing $N$: only for the largest value of $\alpha$ reported in this Figure, namely $\alpha =3$,
we can observe an approach to the expected anomalous scaling for sufficiently large values of $N$.
Moreover, a study of the flux elements $J_{ik}$ 
 performed for the system
 sizes $N=64,$ $128$ and $256$
 suggests that a decomposition similar to  Eq.~(\ref{eq:dec}) holds also for the FPU chain, provided 
 that the diffusive term $N^{-1}g_{ik}^{(\alpha)}$ is 
replaced by an anomalous one of the form $N^{-\frac{3}{5}}g_{ik}^{(\alpha)}$  (data not shown). 
We want to point out that a deeper inspection of this regime is practically demanding, because
of limitations of the available numerical resources. Anyway, we should conclude that  for increasing 
values of $\alpha$ there is a monotonous tendency of $Q$ to recover the expected anomalous scaling 
over a larger and larger domain of $N$.
The results reported in Fig.~\ref{fig:FPUfluxscal} for $\alpha = 2$ (see the red squares)
indicate that this is not the case: $Q$ seems to be independent of $N$. We can exclude
 that this result can be attributed to the parallelization process that sets in for $\alpha < 1$.
Indeed, the inspection  of the current-matrix $\hat J$ reveals that, for $\alpha = 2$,  the relevant energy 
channels are exclusively those around the diagonal, as shown in panel (a) of Fig.~\ref{fig:FMAT}.
At first sight, it seems that in this peculiar case we are facing a sort of ballistic regime, characterisitc
of integrable models, e.g. the harmonic oscillator chain or the Toda lattice ~\cite{hu2000heat,LLP03}.          
We point out that the same observation has been reported very recently by Bagchi~\cite{bagchi2017thermal} for a 
FPU-$\beta$ model with tunable range parameter,  where only the nonlinear quartic term was put in the form of Eq.~(\ref{eq:int_pot}), while the harmonic interaction was maintained in the nearest-neighbor form.  
In Ref.~\cite{bagchi2017thermal}, the apparent ballistic transport for $\alpha=2$ was attributed to the emergence of 
a quasi-integrable dynamics. On the other hand, recovering the same kind of peculiar behavior for
$\alpha=2$ in our  version of the long-range FPU-$\beta$ chain testifies at the robustness of this effect, but
makes doubtful the possibility of associating it to integrability, or quasi-integrability, whatever this means.

A better understanding of this phenomenon can be obtained by comparing the temperature profile of the
stationary state for $\alpha = 2$ with those of the harmonic chain and of the Toda lattice.  For what concerns
the harmonic chain, the temperature profile can be computed analytically \cite{RLL67}
and it is known to exhibit for increasing values of $N$ a flatter and flatter shape far from the boundaries around a temperature
$(T_A + T_B)/2$. In the absence of an analytic estimate, numerics shows that a similar 
shape is obtained for the Toda chain. In the inset of Fig.~\ref{fig:FPUfluxscal} we can observe  that for
model (\ref{eq:intFPU}) with $\alpha = 2$ the temperature profile deviates more from flatness the larger 
the system size $N$. On the basis of this observation, we can just argue that finite size effects are playing
a dominant role in this case. This said, it seems quite hard to provide any convincing argument that
could explain what happens for $\alpha = 2$. Preliminary results (not reported here) for 
long-range models with different kinds of  FPU-like interactions confirm the peculiarity of this case
together with the important role played by finite size effects. 

\section{conclusions}
\label{sec7}

Heat transport in long-range models exhibits quite interesting and peculiar features,
that depend on the nature of the interaction potential. The comparison between the
rotor chain and the FPU-$\beta$ chain confirms this scenario, that was already outlined
in some recent pubblications \cite{avila2015length,Olivares16,bagchi2017thermal}.
In this paper we have shown that in the  non-additive regime corresponding to
$ 0 < \alpha < 1$ the mechanism of heat transport is dominated by the direct, mean-field like
contribution, irrespectively of the kind of the interaction potential.

To rationalize the numerical
results, we have proposed  a decomposition of the steady-state flow, Eq.(\ref{eq:dec}), 
that accounts for the direct 
energy transfer between the heat baths. Such decomposition allows us to conclude that for
$\alpha > 1$ heat transport is dominated by local interaction mechanisms, that are
expected to  reproduce the same scaling properties of the total heat flux $Q$ with the
system size $N$,  observed in the nearest-neighbor limit $\alpha \to \infty$. On the other hand,
finite size effects have been found to have a strong influence in the study of heat
transport in long-range models for any finite value of $\alpha$. This is the main
reason why clear-cut conclusions are very hard to be drawn
just relying upon accessible numerical studies. This is the case also when dealing with the
peculiar situation observed in the FPU-$\beta$ chain for $\alpha = 2$.  In this case 
we are facing quite an unexpected behaviour, that seems to be dominated by a ballistic
transport mechanism, rather than reproducing the anomalous scaling of 
$Q$, typical  of the nearest-neighbour limit. The fact that such a peculiarity of the
case $\alpha =2$ seems to emerge in different versions of  long-range FPU-like models
with an ubiquitous  manifestation of subtle finite-size effects testifies at the interest of
this phenomenon, that is still waiting for a convincing explanation.

\acknowledgements
We thank S. Gupta, A. Politi, A. Dhar and Y. Dubi for illuminating discussions. 

\bibliography{biblio}

\begin{thebibliography}{36}
\expandafter\ifx\csname natexlab\endcsname\relax\def\natexlab#1{#1}\fi
\expandafter\ifx\csname bibnamefont\endcsname\relax
  \def\bibnamefont#1{#1}\fi
\expandafter\ifx\csname bibfnamefont\endcsname\relax
  \def\bibfnamefont#1{#1}\fi
\expandafter\ifx\csname citenamefont\endcsname\relax
  \def\citenamefont#1{#1}\fi
\expandafter\ifx\csname url\endcsname\relax
  \def\url#1{\texttt{#1}}\fi
\expandafter\ifx\csname urlprefix\endcsname\relax\def\urlprefix{URL }\fi
\providecommand{\bibinfo}[2]{#2}
\providecommand{\eprint}[2][]{\url{#2}}

\bibitem[{\citenamefont{Lepri et~al.}(2003{\natexlab{a}})\citenamefont{Lepri,
  Livi, and Politi}}]{LLP03}
\bibinfo{author}{\bibfnamefont{S.}~\bibnamefont{Lepri}},
  \bibinfo{author}{\bibfnamefont{R.}~\bibnamefont{Livi}}, \bibnamefont{and}
  \bibinfo{author}{\bibfnamefont{A.}~\bibnamefont{Politi}},
  \bibinfo{journal}{Phys. Rep.} \textbf{\bibinfo{volume}{377}},
  \bibinfo{pages}{1} (\bibinfo{year}{2003}{\natexlab{a}}).

\bibitem[{\citenamefont{Basile et~al.}(2007)\citenamefont{Basile, Delfini,
  Lepri, Livi, Olla, and Politi}}]{Basile08}
\bibinfo{author}{\bibfnamefont{G.}~\bibnamefont{Basile}},
  \bibinfo{author}{\bibfnamefont{L.}~\bibnamefont{Delfini}},
  \bibinfo{author}{\bibfnamefont{S.}~\bibnamefont{Lepri}},
  \bibinfo{author}{\bibfnamefont{R.}~\bibnamefont{Livi}},
  \bibinfo{author}{\bibfnamefont{S.}~\bibnamefont{Olla}}, \bibnamefont{and}
  \bibinfo{author}{\bibfnamefont{A.}~\bibnamefont{Politi}},
  \bibinfo{journal}{Eur. Phys J.-Special Topics}
  \textbf{\bibinfo{volume}{151}}, \bibinfo{pages}{85} (\bibinfo{year}{2007}).

\bibitem[{\citenamefont{Dhar}(2008)}]{DHARREV}
\bibinfo{author}{\bibfnamefont{A.}~\bibnamefont{Dhar}}, \bibinfo{journal}{Adv.
  Phys.} \textbf{\bibinfo{volume}{57}}, \bibinfo{pages}{457}
  (\bibinfo{year}{2008}).

\bibitem[{\citenamefont{Lepri}(2016)}]{Lepri2016}
\bibinfo{editor}{\bibfnamefont{S.}~\bibnamefont{Lepri}}, ed.,
  \emph{\bibinfo{title}{Thermal transport in low dimensions: from statistical
  physics to nanoscale heat transfer}}, vol. \bibinfo{volume}{921} of
  \emph{\bibinfo{series}{Lect. Notes Phys}}
  (\bibinfo{publisher}{Springer-Verlag, Berlin Heidelberg},
  \bibinfo{year}{2016}).

\bibitem[{\citenamefont{{\'A}vila et~al.}(2015)\citenamefont{{\'A}vila,
  Pereira, and Teixeira}}]{avila2015length}
\bibinfo{author}{\bibfnamefont{R.~R.} \bibnamefont{{\'A}vila}},
  \bibinfo{author}{\bibfnamefont{E.}~\bibnamefont{Pereira}}, \bibnamefont{and}
  \bibinfo{author}{\bibfnamefont{D.~L.} \bibnamefont{Teixeira}},
  \bibinfo{journal}{Physica A: Statistical Mechanics and its Applications}
  \textbf{\bibinfo{volume}{423}}, \bibinfo{pages}{51} (\bibinfo{year}{2015}).

\bibitem[{\citenamefont{{Olivares} and {Anteneodo}}(2016)}]{Olivares16}
\bibinfo{author}{\bibfnamefont{C.}~\bibnamefont{{Olivares}}} \bibnamefont{and}
  \bibinfo{author}{\bibfnamefont{C.}~\bibnamefont{{Anteneodo}}},
  \bibinfo{journal}{\pre} \textbf{\bibinfo{volume}{94}}, \bibinfo{eid}{042117}
  (\bibinfo{year}{2016}), \eprint{1604.07820}.

\bibitem[{\citenamefont{Bagchi}(2017)}]{bagchi2017thermal}
\bibinfo{author}{\bibfnamefont{D.}~\bibnamefont{Bagchi}},
  \bibinfo{journal}{Phys. Rev. E} \textbf{\bibinfo{volume}{95}},
  \bibinfo{pages}{032102} (\bibinfo{year}{2017}).

\bibitem[{\citenamefont{Bouchet et~al.}(2010)\citenamefont{Bouchet, Gupta, and
  Mukamel}}]{Bouchet2010}
\bibinfo{author}{\bibfnamefont{F.}~\bibnamefont{Bouchet}},
  \bibinfo{author}{\bibfnamefont{S.}~\bibnamefont{Gupta}}, \bibnamefont{and}
  \bibinfo{author}{\bibfnamefont{D.}~\bibnamefont{Mukamel}},
  \bibinfo{journal}{Physica A: Statistical Mechanics and its Applications}
  \textbf{\bibinfo{volume}{389}}, \bibinfo{pages}{4389} (\bibinfo{year}{2010}).

\bibitem[{\citenamefont{Campa et~al.}(2009)\citenamefont{Campa, Dauxois, and
  Ruffo}}]{RuffoRev}
\bibinfo{author}{\bibfnamefont{A.}~\bibnamefont{Campa}},
  \bibinfo{author}{\bibfnamefont{T.}~\bibnamefont{Dauxois}}, \bibnamefont{and}
  \bibinfo{author}{\bibfnamefont{S.}~\bibnamefont{Ruffo}},
  \bibinfo{journal}{Phys. Rep.} \textbf{\bibinfo{volume}{480}},
  \bibinfo{pages}{57} (\bibinfo{year}{2009}).

\bibitem[{\citenamefont{Campa et~al.}(2014)\citenamefont{Campa, Dauxois,
  Fanelli, and Ruffo}}]{Campa2014}
\bibinfo{author}{\bibfnamefont{A.}~\bibnamefont{Campa}},
  \bibinfo{author}{\bibfnamefont{T.}~\bibnamefont{Dauxois}},
  \bibinfo{author}{\bibfnamefont{D.}~\bibnamefont{Fanelli}}, \bibnamefont{and}
  \bibinfo{author}{\bibfnamefont{S.}~\bibnamefont{Ruffo}},
  \emph{\bibinfo{title}{Physics of long-range interacting systems}}
  (\bibinfo{publisher}{OUP Oxford}, \bibinfo{year}{2014}).

\bibitem[{\citenamefont{de~Buyl et~al.}(2013)\citenamefont{de~Buyl, De~Ninno,
  Fanelli, Nardini, Patelli, Piazza, and Yamaguchi}}]{deBuyl2013}
\bibinfo{author}{\bibfnamefont{P.}~\bibnamefont{de~Buyl}},
  \bibinfo{author}{\bibfnamefont{G.}~\bibnamefont{De~Ninno}},
  \bibinfo{author}{\bibfnamefont{D.}~\bibnamefont{Fanelli}},
  \bibinfo{author}{\bibfnamefont{C.}~\bibnamefont{Nardini}},
  \bibinfo{author}{\bibfnamefont{A.}~\bibnamefont{Patelli}},
  \bibinfo{author}{\bibfnamefont{F.}~\bibnamefont{Piazza}}, \bibnamefont{and}
  \bibinfo{author}{\bibfnamefont{Y.~Y.} \bibnamefont{Yamaguchi}},
  \bibinfo{journal}{Phys. Rev. E} \textbf{\bibinfo{volume}{87}},
  \bibinfo{pages}{042110} (\bibinfo{year}{2013}).

\bibitem[{\citenamefont{{Teles} et~al.}(2015)\citenamefont{{Teles}, {Gupta},
  {Di Cintio}, and {Casetti}}}]{2015PhRvE..92b0101T}
\bibinfo{author}{\bibfnamefont{T.~N.} \bibnamefont{{Teles}}},
  \bibinfo{author}{\bibfnamefont{S.}~\bibnamefont{{Gupta}}},
  \bibinfo{author}{\bibfnamefont{P.}~\bibnamefont{{Di Cintio}}},
  \bibnamefont{and}
  \bibinfo{author}{\bibfnamefont{L.}~\bibnamefont{{Casetti}}},
  \bibinfo{journal}{\pre} \textbf{\bibinfo{volume}{92}}, \bibinfo{eid}{020101}
  (\bibinfo{year}{2015}), \eprint{1502.04051}.

\bibitem[{\citenamefont{Gupta and Casetti}(2016)}]{gupta2016surprises}
\bibinfo{author}{\bibfnamefont{S.}~\bibnamefont{Gupta}} \bibnamefont{and}
  \bibinfo{author}{\bibfnamefont{L.}~\bibnamefont{Casetti}},
  \bibinfo{journal}{New Journal of Physics} \textbf{\bibinfo{volume}{18}},
  \bibinfo{pages}{103051} (\bibinfo{year}{2016}).

\bibitem[{\citenamefont{Torcini and Lepri}(1997)}]{Torcini1997}
\bibinfo{author}{\bibfnamefont{A.}~\bibnamefont{Torcini}} \bibnamefont{and}
  \bibinfo{author}{\bibfnamefont{S.}~\bibnamefont{Lepri}},
  \bibinfo{journal}{Phys. Rev. E} \textbf{\bibinfo{volume}{55}},
  \bibinfo{pages}{R3805} (\bibinfo{year}{1997}).

\bibitem[{\citenamefont{M{\'e}tivier et~al.}(2014)\citenamefont{M{\'e}tivier,
  Bachelard, and Kastner}}]{Metivier2014}
\bibinfo{author}{\bibfnamefont{D.}~\bibnamefont{M{\'e}tivier}},
  \bibinfo{author}{\bibfnamefont{R.}~\bibnamefont{Bachelard}},
  \bibnamefont{and} \bibinfo{author}{\bibfnamefont{M.}~\bibnamefont{Kastner}},
  \bibinfo{journal}{Phys. Rev. Lett.} \textbf{\bibinfo{volume}{112}},
  \bibinfo{pages}{210601} (\bibinfo{year}{2014}).

\bibitem[{\citenamefont{Giardin{\`a} et~al.}(2000)\citenamefont{Giardin{\`a},
  Livi, Politi, and Vassalli}}]{Giardina99}
\bibinfo{author}{\bibfnamefont{C.}~\bibnamefont{Giardin{\`a}}},
  \bibinfo{author}{\bibfnamefont{R.}~\bibnamefont{Livi}},
  \bibinfo{author}{\bibfnamefont{A.}~\bibnamefont{Politi}}, \bibnamefont{and}
  \bibinfo{author}{\bibfnamefont{M.}~\bibnamefont{Vassalli}},
  \bibinfo{journal}{Phys. Rev. Lett.} \textbf{\bibinfo{volume}{84}},
  \bibinfo{pages}{2144} (\bibinfo{year}{2000}), ISSN \bibinfo{issn}{0031-9007}.

\bibitem[{\citenamefont{Gendelman and Savin}(2000)}]{gendelman2000}
\bibinfo{author}{\bibfnamefont{O.}~\bibnamefont{Gendelman}} \bibnamefont{and}
  \bibinfo{author}{\bibfnamefont{A.}~\bibnamefont{Savin}},
  \bibinfo{journal}{Phys. Rev. Lett.} \textbf{\bibinfo{volume}{84}},
  \bibinfo{pages}{2381} (\bibinfo{year}{2000}).

\bibitem[{\citenamefont{Rieder et~al.}(1967)\citenamefont{Rieder, Lebowitz, and
  Lieb}}]{RLL67}
\bibinfo{author}{\bibfnamefont{Z.}~\bibnamefont{Rieder}},
  \bibinfo{author}{\bibfnamefont{J.~L.} \bibnamefont{Lebowitz}},
  \bibnamefont{and} \bibinfo{author}{\bibfnamefont{E.}~\bibnamefont{Lieb}},
  \bibinfo{journal}{J. Math. Phys.} \textbf{\bibinfo{volume}{8}},
  \bibinfo{pages}{1073} (\bibinfo{year}{1967}).

\bibitem[{\citenamefont{Toda}(1979)}]{Toda79}
\bibinfo{author}{\bibfnamefont{M.}~\bibnamefont{Toda}}, \bibinfo{journal}{Phys.
  Scr.} \textbf{\bibinfo{volume}{20}}, \bibinfo{pages}{424}
  (\bibinfo{year}{1979}), ISSN \bibinfo{issn}{0281-1847}.

\bibitem[{\citenamefont{Kundu and Dhar}(2016)}]{Kundu2016}
\bibinfo{author}{\bibfnamefont{A.}~\bibnamefont{Kundu}} \bibnamefont{and}
  \bibinfo{author}{\bibfnamefont{A.}~\bibnamefont{Dhar}},
  \bibinfo{journal}{Phys. Rev. E} \textbf{\bibinfo{volume}{94}},
  \bibinfo{pages}{062130} (\bibinfo{year}{2016}).

\bibitem[{\citenamefont{Tamarit and Anteneodo}(2000)}]{Tamarit2000}
\bibinfo{author}{\bibfnamefont{F.}~\bibnamefont{Tamarit}} \bibnamefont{and}
  \bibinfo{author}{\bibfnamefont{C.}~\bibnamefont{Anteneodo}},
  \bibinfo{journal}{Phys. Rev. Lett.} \textbf{\bibinfo{volume}{84}},
  \bibinfo{pages}{208} (\bibinfo{year}{2000}).

\bibitem[{\citenamefont{{Van Den Berg} et~al.}(2010)\citenamefont{{Van Den
  Berg}, Fanelli, and Leoncini}}]{vanderberg2010}
\bibinfo{author}{\bibfnamefont{T.~L.} \bibnamefont{{Van Den Berg}}},
  \bibinfo{author}{\bibfnamefont{D.}~\bibnamefont{Fanelli}}, \bibnamefont{and}
  \bibinfo{author}{\bibfnamefont{X.}~\bibnamefont{Leoncini}},
  \bibinfo{journal}{EPL (Europhysics Letters)} \textbf{\bibinfo{volume}{89}},
  \bibinfo{pages}{50010} (\bibinfo{year}{2010}).

\bibitem[{\citenamefont{Yang and Hu}(2005)}]{Yang2005}
\bibinfo{author}{\bibfnamefont{L.}~\bibnamefont{Yang}} \bibnamefont{and}
  \bibinfo{author}{\bibfnamefont{B.}~\bibnamefont{Hu}}, \bibinfo{journal}{Phys.
  Rev. Lett.} \textbf{\bibinfo{volume}{94}}, \bibinfo{pages}{219404}
  (\bibinfo{year}{2005}).

\bibitem[{\citenamefont{Iubini et~al.}(2016)\citenamefont{Iubini, Lepri, Livi,
  and Politi}}]{Iubini2016}
\bibinfo{author}{\bibfnamefont{S.}~\bibnamefont{Iubini}},
  \bibinfo{author}{\bibfnamefont{S.}~\bibnamefont{Lepri}},
  \bibinfo{author}{\bibfnamefont{R.}~\bibnamefont{Livi}}, \bibnamefont{and}
  \bibinfo{author}{\bibfnamefont{A.}~\bibnamefont{Politi}},
  \bibinfo{journal}{New J. Phys.} \textbf{\bibinfo{volume}{18}},
  \bibinfo{pages}{083023} (\bibinfo{year}{2016}).

\bibitem[{\citenamefont{Lepri et~al.}(1997)\citenamefont{Lepri, Livi, and
  Politi}}]{LLP97}
\bibinfo{author}{\bibfnamefont{S.}~\bibnamefont{Lepri}},
  \bibinfo{author}{\bibfnamefont{R.}~\bibnamefont{Livi}}, \bibnamefont{and}
  \bibinfo{author}{\bibfnamefont{A.}~\bibnamefont{Politi}},
  \bibinfo{journal}{Phys. Rev. Lett.} \textbf{\bibinfo{volume}{78}},
  \bibinfo{pages}{1896} (\bibinfo{year}{1997}), ISSN \bibinfo{issn}{0031-9007}.

\bibitem[{\citenamefont{Lepri et~al.}(2005)\citenamefont{Lepri, Livi, and
  Politi}}]{Lepri05}
\bibinfo{author}{\bibfnamefont{S.}~\bibnamefont{Lepri}},
  \bibinfo{author}{\bibfnamefont{R.}~\bibnamefont{Livi}}, \bibnamefont{and}
  \bibinfo{author}{\bibfnamefont{A.}~\bibnamefont{Politi}},
  \bibinfo{journal}{Chaos} \textbf{\bibinfo{volume}{15}},
  \bibinfo{pages}{015118} (\bibinfo{year}{2005}), ISSN
  \bibinfo{issn}{1054-1500}.

\bibitem[{\citenamefont{Wang and Wang}(2011)}]{Wang2011}
\bibinfo{author}{\bibfnamefont{L.}~\bibnamefont{Wang}} \bibnamefont{and}
  \bibinfo{author}{\bibfnamefont{T.}~\bibnamefont{Wang}}, \bibinfo{journal}{EPL
  (Europhysics Letters)} \textbf{\bibinfo{volume}{93}}, \bibinfo{pages}{54002}
  (\bibinfo{year}{2011}).

\bibitem[{\citenamefont{Miloshevich et~al.}(2015)\citenamefont{Miloshevich,
  Nguenang, Dauxois, Khomeriki, and Ruffo}}]{Miloshevich2015}
\bibinfo{author}{\bibfnamefont{G.}~\bibnamefont{Miloshevich}},
  \bibinfo{author}{\bibfnamefont{J.-P.} \bibnamefont{Nguenang}},
  \bibinfo{author}{\bibfnamefont{T.}~\bibnamefont{Dauxois}},
  \bibinfo{author}{\bibfnamefont{R.}~\bibnamefont{Khomeriki}},
  \bibnamefont{and} \bibinfo{author}{\bibfnamefont{S.}~\bibnamefont{Ruffo}},
  \bibinfo{journal}{Phys. Rev. E} \textbf{\bibinfo{volume}{91}},
  \bibinfo{pages}{032927} (\bibinfo{year}{2015}).

\bibitem[{\citenamefont{Miloshevich et~al.}(2017)\citenamefont{Miloshevich,
  Nguenang, Dauxois, Khomeriki, and Ruffo}}]{Miloshevich2017}
\bibinfo{author}{\bibfnamefont{G.}~\bibnamefont{Miloshevich}},
  \bibinfo{author}{\bibfnamefont{J.~P.} \bibnamefont{Nguenang}},
  \bibinfo{author}{\bibfnamefont{T.}~\bibnamefont{Dauxois}},
  \bibinfo{author}{\bibfnamefont{R.}~\bibnamefont{Khomeriki}},
  \bibnamefont{and} \bibinfo{author}{\bibfnamefont{S.}~\bibnamefont{Ruffo}},
  \bibinfo{journal}{J. Phys. A: Math. Theor.} \textbf{\bibinfo{volume}{50}},
  \bibinfo{pages}{12LT02} (\bibinfo{year}{2017}).

\bibitem[{\citenamefont{McLachlan and Atela}(1992)}]{mclachlan1992accuracy}
\bibinfo{author}{\bibfnamefont{R.~I.} \bibnamefont{McLachlan}}
  \bibnamefont{and} \bibinfo{author}{\bibfnamefont{P.}~\bibnamefont{Atela}},
  \bibinfo{journal}{Nonlinearity} \textbf{\bibinfo{volume}{5}},
  \bibinfo{pages}{541} (\bibinfo{year}{1992}).

\bibitem[{\citenamefont{Liu and Li}(2007)}]{liu2007heat}
\bibinfo{author}{\bibfnamefont{Z.}~\bibnamefont{Liu}} \bibnamefont{and}
  \bibinfo{author}{\bibfnamefont{B.}~\bibnamefont{Li}}, \bibinfo{journal}{Phys.
  Rev. E} \textbf{\bibinfo{volume}{76}}, \bibinfo{pages}{051118}
  (\bibinfo{year}{2007}).

\bibitem[{\citenamefont{Liu et~al.}(2010)\citenamefont{Liu, Wu, Yang, Gupte,
  and Li}}]{liu2010heat}
\bibinfo{author}{\bibfnamefont{Z.}~\bibnamefont{Liu}},
  \bibinfo{author}{\bibfnamefont{X.}~\bibnamefont{Wu}},
  \bibinfo{author}{\bibfnamefont{H.}~\bibnamefont{Yang}},
  \bibinfo{author}{\bibfnamefont{N.}~\bibnamefont{Gupte}}, \bibnamefont{and}
  \bibinfo{author}{\bibfnamefont{B.}~\bibnamefont{Li}}, \bibinfo{journal}{New
  J. Phys.} \textbf{\bibinfo{volume}{12}}, \bibinfo{pages}{023016}
  (\bibinfo{year}{2010}).

\bibitem[{\citenamefont{Lepri et~al.}(2003{\natexlab{b}})\citenamefont{Lepri,
  Livi, and Politi}}]{LLP2003}
\bibinfo{author}{\bibfnamefont{S.}~\bibnamefont{Lepri}},
  \bibinfo{author}{\bibfnamefont{R.}~\bibnamefont{Livi}}, \bibnamefont{and}
  \bibinfo{author}{\bibfnamefont{A.}~\bibnamefont{Politi}},
  \bibinfo{journal}{Phys. Rev. E} \textbf{\bibinfo{volume}{68}},
  \bibinfo{pages}{067102} (\bibinfo{year}{2003}{\natexlab{b}}).

\bibitem[{\citenamefont{Pereverzev}(2003)}]{Pereverzev2003}
\bibinfo{author}{\bibfnamefont{A.}~\bibnamefont{Pereverzev}},
  \bibinfo{journal}{Phys. Rev. E} \textbf{\bibinfo{volume}{68}},
  \bibinfo{pages}{056124} (\bibinfo{year}{2003}).

\bibitem[{\citenamefont{Lukkarinen and Spohn}(2008)}]{Lukkarinen2008}
\bibinfo{author}{\bibfnamefont{J.}~\bibnamefont{Lukkarinen}} \bibnamefont{and}
  \bibinfo{author}{\bibfnamefont{H.}~\bibnamefont{Spohn}},
  \bibinfo{journal}{Communications on Pure and Applied Mathematics}
  \textbf{\bibinfo{volume}{61}}, \bibinfo{pages}{1753} (\bibinfo{year}{2008}).

\bibitem[{\citenamefont{Hu et~al.}(2000)\citenamefont{Hu, Li, and
  Zhao}}]{hu2000heat}
\bibinfo{author}{\bibfnamefont{B.}~\bibnamefont{Hu}},
  \bibinfo{author}{\bibfnamefont{B.}~\bibnamefont{Li}}, \bibnamefont{and}
  \bibinfo{author}{\bibfnamefont{H.}~\bibnamefont{Zhao}},
  \bibinfo{journal}{Physical Review E} \textbf{\bibinfo{volume}{61}},
  \bibinfo{pages}{3828} (\bibinfo{year}{2000}).

\end{thebibliography}

\bibliographystyle{apsrev}

\end{document}